\documentclass [a4paper,10pt,english]{article} 
\usepackage[top=1.2in, bottom=0.8in, left=.6in, right=.6in]{geometry}
\usepackage{amsmath} 
\usepackage[utf8]{inputenc} 
\usepackage[bookmarks=false, colorlinks=false,unicode]{hyperref}
\usepackage[T1]{fontenc}
\usepackage{hyperref} 
\usepackage{graphicx} 
\usepackage{listings} 
\usepackage{subfigure}                  
\usepackage{physics}
\usepackage{amsmath, amsthm, amssymb, amsfonts}
\usepackage{authblk}
\usepackage{blindtext}
\usepackage{xcolor}

\begin{document}

\title{Detection of the Quantum Illumination Measurement}
\author[1,2]{Michal Krelina\thanks{michal.krelina@quantumphi.com}}
\date{June 25, 2019}
\affil[1]{
Quantum Phi s.r.o., Bryksova 944/27, 18000 Prague, Czech Republic\\
}
\affil[2]{
Czech Technical University in Prague, Faculty of Nuclear Sciences and Physical Engineering, Brehova 7, 11519 Prague, Czech Republic
}
\maketitle

\begin{abstract}
In this report, we discuss possibilities to detect a signal at the target from the quantum illumination protocol, that could serve as a quantum radar.
We assume a simple universal detecting schema on the target and study if it is possible to discover the quantum illumination measurement and in what conditions considering the microwave regime.
Assuming many simplifications, we found that the possibility or the advantage of the detection of the quantum illumination measurement strongly depends on the realization of the quantum illumination protocol.
\end{abstract}

%
%

%
%
\section{Introduction}

Quantum illumination (QI) is a quantum protocol for detecting and imaging objects in a noisy and lossy environment utilizing quantum entanglement \cite{Lloyd:2008lqd} and as such is suitable for quantum radar applications.
Within this protocol, two entangled beams denoted as an ancilla (idler) and as a signal, are created. The ancilla beam is retained in the system, and the signal beam is sent to the region where the target may present.
Then, the joint measurement of ancilla and the reflected signal photons is performed.
Quantum illumination is in the spotlight of the current research because it preserves the substantial advantage in the error probability in comparison with classical protocols of the same average photon number.
The advantage of quantum illuminations is carried on in such an environment where no entanglement survives at the detector due to strong noise and losses.

Quantum illumination has been intensively studied, and the QI protocol was extended from single photon \cite{Lloyd:2008lqd} over multiphoton \cite{Shapiro:2009qiv} up to quantum Gaussian states \cite{Tan:2008qig}.
Quantum illumination was for the first time experimentally realized in \cite{Lopaeva:2013urt} and then fully confirmed in \cite{Zhang:2015kjr}. QI is therefore an ideal protocol for quantum sensing, see e.g., \cite{Lloyd:2008lqd,Shapiro:2009qiv,Tan:2008qig,Guha:2009iie,Invernizzi:2011fwa,Chang:2018nfg}.

Quantum radar (QR) \cite{Lanzagorta:QuantumRadar} is an emerging quantum technology that has a potential revolutionize the radar technology with a significant consequence, especially in the defense domain.
Theoretically, quantum radar could offer target (even stealth target) detection, tracking, and illumination in a noisy environment and jamming resistance. Moreover, the quantum radar itself would be invisible for the nowadays electronic warfare systems.
For quantum radar is a crucial application of the QI protocol for microwave photons regime \cite{Barzanjeh:2015zpa,England:2019nfj}.

\begin{figure}[htb]
\centering
\includegraphics[width=0.5\textwidth]{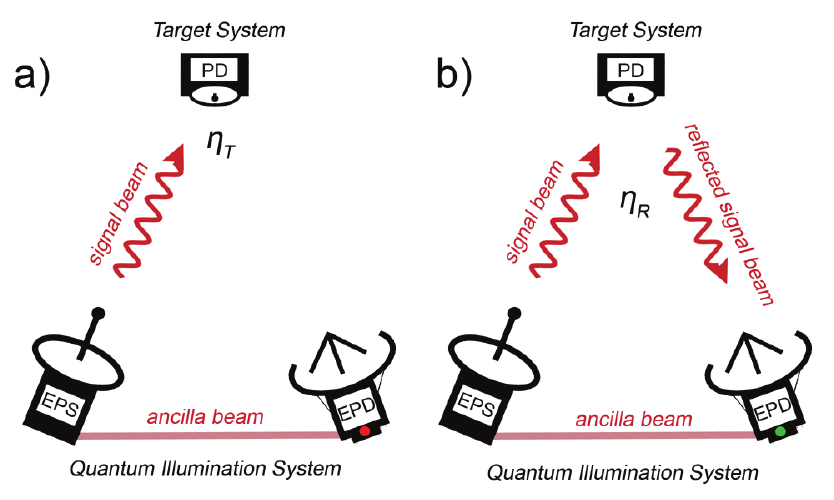}
\caption{a) Detection of the signal beam at the target system (TS). b) Detection of the signal beam reflected from the target in the quantum illumination system (QIS). EPS stands for entangled photon source, PD - photon detector, EPD - entangled photons detector.}
\label{fig:qrwr}
\end{figure}

Since one party uses quantum radar, it is a legitime question whether the other party (target) can detect the acting of quantum radar.
The study of the possibility of detection of the quantum radar, so-called quantum radar warning receiver (QRWR), is the motivation of this report.
For this purpose, we define the quantum illumination system (QIS) as a quantum radar, see Fig.~\ref{fig:qrwr} down, and the target detection system (TS), see Fig.~\ref{fig:qrwr} up, which tries to detect the signal beam from the QIS.
The advantage of quantum illumination is in the correlation of the ancilla and signal beams where the ancilla beam is kept in the QI system and is not available to the target, or, generally, any measurement outside the QIS. Therefore, we lost this significant advantage of strong quantum correlations in the detection at the target. On the other side, the advantage at the target side is lower signal beam loss, $\eta_T$, which represents the losses during the signal beam propagation to the target and the detection efficiency on the target. The loss at the quantum illumination system, $\eta_{R}$, includes not only losses due to the propagation to/from the target and detector efficiency but also the probability of reflection from the target back to the detector of the quantum illumination system, i.e., $\eta_T \gg \eta_{R}$.

This report is organized as follows.
In Sec.~\ref{sec:dete}, we introduce the detection of the signal beam at the target, and we discuss the background that is the key to the feasibility of the QRWR.
Sec.~\ref{sec:meas} shortly introduce the QI protocol and the error probability.
In Sec.~\ref{sec:res}, we present the numerical results and comments.
Finally, the conclusions are in Sec.~\ref{sec:concl}.

%
%
\section{Detection on the target}\label{sec:dete}
The detection of a signal in a lossy and noisy environment is usually a subject of quantum hypothesis testing (or quantum discrimination) \cite{Pirandola:2018dfr}. In most cases, the main idea is to use some correlations, in the best case, the correlations coming from the quantum entanglement, e.g., the quantum illumination, ghost imaging, etc. For basic overview, see for example \cite{Genovese:2016rjb,Meda:2017uir,Pirandola:2018dfr}.

However, in our studied case, we can potentially access only to the signal beam from the quantum illumination system, whereas the idler beam is kept inside the QIS and is not accessed outside this system. Moreover, if we are interested in the application of QI as a quantum radar in the microwave regime, the strong background is expected, i.e., the number of detected background photons is much higher than the number of signal photons, $N_B \gg N_S$.

On the other hand, the advantage of the detection on the target lies in the lower loss in comparison with the loss of the signal beam for the QI system.
In principle, we have a photon loss coming from the attenuation during the propagation towards the target, $\eta_T$. On the other hand, the QI measurement accumulates other losses such as the attenuation during the propagation back to the QI system and the reflectivity of the target exactly back to the detector of QIS, $\eta_R$. Especially, the latter one - reflectivity can be assumed at large distances orders of magnitude smaller.

The task of detection of single photons signals in noisy environment is nowadays a common task of the free-space quantum communication and quantum key distribution (QKD) in optical regime, \cite{buttler_daylight_2000,nordholt_present_2002,erven_entangled_2008,bourgoin_comprehensive_2013}.
The free-space QKD is an example where the single photons carry the quantum information, and it is highly desirable to detect a maximum of them. For this purpose, different methods of how to suppress the background photons are developed and applied \cite{er-long_background_2005}.

Detection of the signal in the quantum radar corresponds to the measurement of Bob where Alice gave him all possible information about the signal. That allows us to apply the frequency (wavelength) filter correctly, and theoretically, the spatial filter and the time-gate filter \cite{er-long_background_2005} if we will approximately know the position and the direction of the target.
On the other side, the detection on the target corresponds to the situation of Bob, where is no information from Alice that makes the detection much complicated. Using the intelligence information about the quantum radar's performance and by suitable design, it could be possible partly apply the frequency and spatial filters (e.g., the target can have more detectors that scan different spatial sector with different spatial filters or a dynamic spatial filter that scan each spatial sector for some amount of time). The significant difference is that we assume the microwave regime instead of optical regime usually used for free-space quantum communication. This dramatically increases the demand to determine the background, more in the following Sec.~\ref{sec:back}.

The measurement itself is then reduced to the detection of background photons, $N_B$, and signal photons $\eta_T N_S$, where the general efficiency $\eta_T$ consist of
$$
\eta_T = \eta_{A} \cdot \eta_{DE} \cdot \eta_{DA},
$$
where $\eta_{A}$ are losses due to the atmosphere attenuation (scattering and absorption), $\eta_{DE}$ is the detector efficiency and $\eta_{DA}$ is a probability that the signal photon from the effective area of the target hits the area of the detector assuming only photons that hits the target.

Because of expected $N_S \ll N_B$, the proposed measurement on the target is the discrimination between the situation when the background only is presented, $N_T = N_B$, and the background plus the signal, $N_T = N_B + \eta_T N_S$ which lead to the signal-to-noise ratio (SNR)
\begin{equation}\label{eq:SNR-DR}
  \textrm{SNR}_T = \frac{\sqrt{K_T} \eta_T N_S }{\sqrt{\delta^2 N_S + 2 \delta^2 N_B  }}
  \approx \frac{\sqrt{K} \eta_T N_S }{\sqrt{2\delta^2 N_B }}  ,
\end{equation}
where $K$ is the number of measurement. For our purpose, we will call it the QRWR protocol.

\subsection{Background}\label{sec:back}

From Eq.~\eqref{eq:SNR-DR}, it is clear that the SNR strongly depends on the variance of the background photons.
In the case of free-space QKD, it is required $SNR > 1$, i.e. approx. $N_S > N_B$. This is usually reached by applying the frequency, spatial, and time-get filters mentioned above.

The free-space detection in the microwave regime is in a substantial disadvantage in comparison with the optical regime. Assuming the sun as the primary source of background photons that can be modeled as black body radiation of the temperature approx. $6000^\circ$C, we can quickly found that less energetic microwave photons are $\sim 10^5$ more than the high energy optical photons.

The microwave regime is actively also investigated in relation to the QKD, see e.g., \cite{weedbrook_quantum_2010,weedbrook_continuous-variable_2012}. However, the general conclusion is that Bob would not be able to distinguish the photons originated from Alice or background; both are indistinguishable. In other words, the microwave regime is not suitable for exchange of quantum information. However, in our case, we simply count the number of photons, which is experimentally very challenging. In the best case, we only can say that "something" above the background is there, and by using the spatial filters, we can approximately say in what direction.

Assuming $N_S<1$ we can say that it will be impossible even with all filters to reach $SNR > 1$. However, in our case we don't rely on such strong requirement since no every single photon doesn't carry unique information as in the QKD, we can increase the SNR by repeated measurement, $K$.

On the other hand, the large amount of measurement $K$ leads to higher time consumption during which the background can significantly fluctuate because of different sources. Except the direct solar photons from the sun, around are many objects that can reflect these solar photons \cite{er-long_background_2005}, there is also the cosmic microwave background, and even a large amount of the microwave communication transmissions worldwide \cite{RFdemands,brown_history_1984}.

Therefore, it can be very difficult to reliably estimate the variance of the background photons, especially in a long time span. Likely, it may be impossible to detect the quantum radar. However, for the final conclusions, more detailed studies on microwave photon background are necessary.

In the rest of the text, we will assume only a well-controlled thermal noise constant in time to study whether the considered advantage of $\eta_T \gg \eta_R$ is considerable in principle.

%
%
\section{Quantum Illumination Measurement}\label{sec:meas}

In \cite{Lloyd:2008lqd}, the quantum illumination protocol was presented, and the theoretical limit for detecting was discussed. However, no particular detection scheme was introduced.

In this work, we will come out of two possible realizations of QI represented by the single-photon (SP) schema presented in \cite{Lopaeva:2013urt,Lopaeva:2014tvv} and the second schema
using the Gaussian states (GS), particularly the two-mode squeezed state obtained from continuous-wave spontaneous parametric downconversion (SPDC) \cite{Tan:2008qig}, more e.g. in \cite{Zhang:2015kjr} or \cite{Barzanjeh:2015zpa,Guha:2009iie}.

The first, the SP schema is more accessible to realize and more corresponds to the real-world requirements on price-performance ratio. However, it does not utilize the full potential of the entanglement as the GS schema based on the homodyne-detection receiver \cite{Guha:2009iie}. The main disadvantage of the GS schema using the homodyne detection is the requirement on idler beam storage, which storage time is appropriate to the target distance.

For the SP schema, the signal-to-noise ratio (SNR) of the QI measurement is based on the covariance measurement between the total photon numbers $N_1$ and $N_2$ of correlated detectors (one measure signal+noise and one ancilla beam, respectively)
\begin{equation}\label{eq:SNR}
  \textrm{SNR}_{SP} \approx \frac{\sqrt{K_R}  \sqrt{\eta_R \eta_A N_S } }{\sqrt{2\delta^2 N_B }}
\end{equation}
where $\eta_R$  and $\eta_A$ are efficiency for the signal and ancilla beam, respective.

Homodyne-type detection in \cite{Barzanjeh:2015zpa,Guha:2009iie} is based on the phase-conujugation of signal and idler beam and the detection in the balanced difference detector, where
the SNR approximately reads
\begin{equation}\label{eq:SNR-GS}
SNR_{GS} \approx \frac{K_R \eta_R \eta_A N_S}{2\sqrt{\delta^2 N_B }}.
\end{equation}

For both schema, we can assume
\begin{eqnarray}
  \eta_R &=& \eta_{A}^2 \cdot \eta_{X} \cdot \eta_{DE}, \\
  \eta_A &=& \eta_{IE} \cdot \eta_{DE},
\end{eqnarray}
where $\eta_{IE}$ is the efficiency of the ancilla photon system, and  $\eta_{X}$ is a probability that the signal photon reflect exactly back to the QI detector. The square of $\eta_{A}$ represents the attenuation on the way to and back from the target that can be assumed the same.

From SNR, for equally-likely hypotheses, we can get the error detection probability as
\begin{equation}\label{eq:PerrSNR}
  P_{err} = \frac{1}{2}\textrm{erfc}\left( \sqrt{\frac{\textrm{SNR}}{8}} \right).
\end{equation}

The way of how we will compare the performance of the detection of quantum radar and target is a ratio
\begin{equation}\label{eq:RM}
  R_M = \frac{n_T}{n_R} = \frac{K_T}{K_R},
\end{equation}
i.e. the number of photons, or equivalently, the number of measurement, to reach the same SNR. Or more precisely, how many measurement we need to reach the same error detection probability \eqref{eq:PerrSNR}.

The ratio $R_M$ says how many times more photons we need to detect on the target system in comparison with the QI system. We highlight three regimes, first regime, $R_M < 1$, means that the presence of some QI system will be detected on the target system before the QI system discover the target.
The second region, $R_M > 1$, corresponds to the situation when the presence of QI measurement will discover the target before the target notices the QIS presence.
The last region is for $R_M \gg 1$, where it is very difficult or practically impossible to detect the QI measurement at the target.

\section{Results and discussion}\label{sec:res}
First, we will study the $R_M$ ratio, where we assume the detection probability equal to $2\sigma$ (95.45\%) for the detection in the target system as well as in the QR system. Note, we assume the same performance and working cycle for detectors at the target as well as for the QI detector.

\begin{figure}[htb]
\centering
\includegraphics[width=0.5\textwidth]{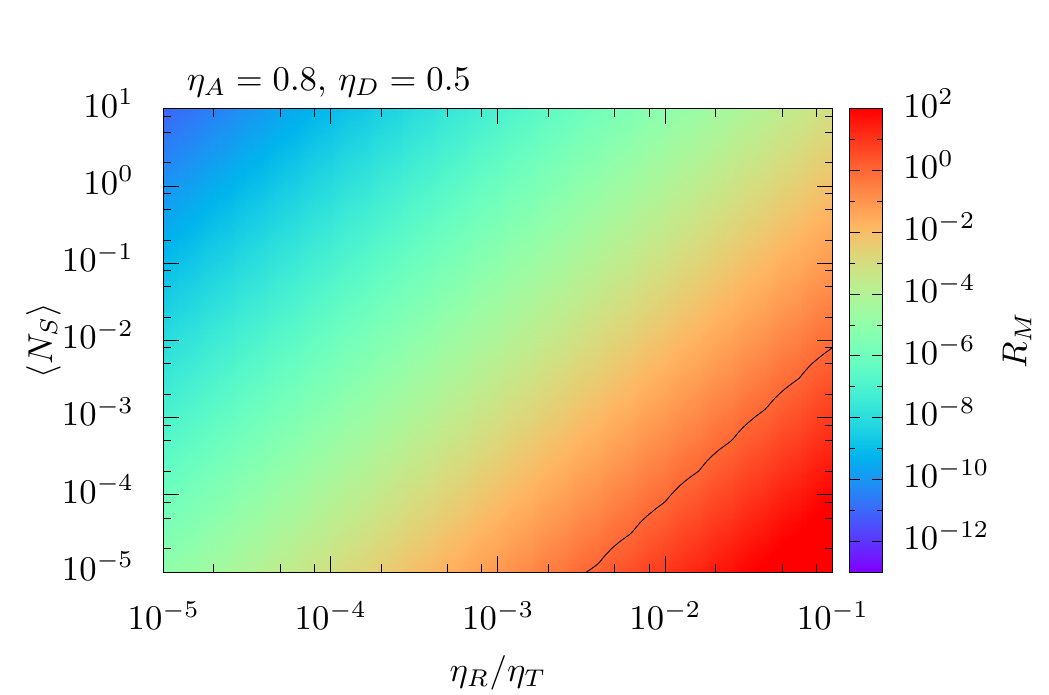}
\caption{Ratio $R_M$ as a function of the number of signal photons $\langle N_S \rangle$ and the ratio $\eta_R/ \eta_T$ for $\eta_T = 0.0001, M=100$. The black line visualize situation for $R_M=1$.}
\label{fig:SP}
\end{figure}

The first result, calculated according \eqref{eq:SNR-DR} and \cite{Lopaeva:2013urt,Lopaeva:2014tvv} and presented in Fig.~\ref{fig:SP}, and shows the ratio $R_M$ for the SP realization of QI as a function of the total number of signal photons $\langle N_S \rangle = M\mu$ and the ratio $\eta_R/ \eta_T$, where $\mu$ is the number of photons in one mode and $M$ is the number of modes (for more detail see \cite{Lopaeva:2013urt,Lopaeva:2014tvv}).
For the calculation in Fig.~\ref{fig:SP} we used the following values, $\mu=0.00001, \eta_T = 0.0001, \eta_A=0.8$, and $M=100$. The black line in Fig.~\ref{fig:SP} visualizes the situation for $R_M=1$, i.e., both, the target system and the QI system, need the same amount of measurement for the required detection probability.

Comparing the SP detection schema in the QI system and the detection schema at the target, we can find that QI would have the advantage only for $\langle N_S \rangle = M\mu < 1$ and $\eta_R/\eta_T \rightarrow 1$. On the contrary, the target could detect the QI measurement in advance and does some counteraction to avoid its revealing.

Assuming the same background at the QI system and the target system, we found negligible dependence on the number of background photons that can be neglected. In the case of the different backgrounds, we have approximately $R_M \rightarrow R_M \frac{\langle N_B^{target} \rangle}{\langle N_B^{radar} \rangle}$, where $\langle N_B \rangle$ is the total number of background photons in all modes.

\begin{figure}[htb]
\centering
\includegraphics[width=0.5\textwidth]{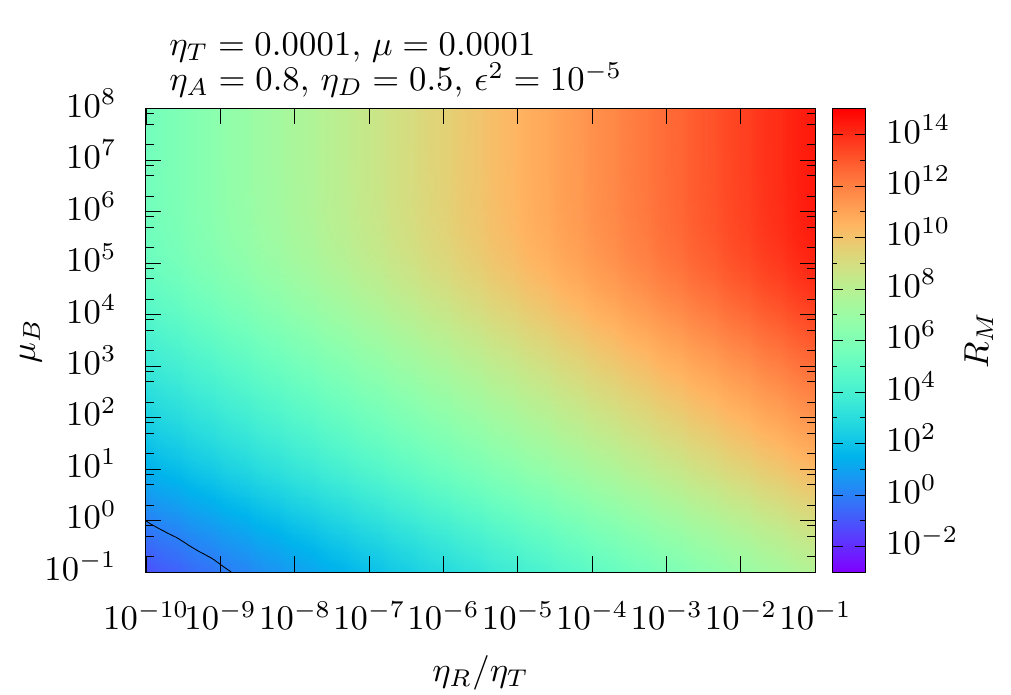}
\caption{Ratio $R_M$ as a function of the number of background photons $N_B$ and the ratio $\eta_R/ \eta_T$ for $\eta_T = 0.0001$. The black line visualize situation for $R_M=1$.}
\label{fig:GS}
\end{figure}

In Fig.~\ref{fig:GS} we show the $R_M$ ratio for the GS realisation of QI as function of the ratio $\eta_R/ \eta_T$ and number of background photons, calculated according \eqref{eq:SNR-DR} and \cite{Guha:2009iie,Tan:2008qig}. This figure demonstrate the significant advantage of the QI system fully utilizing the entanglement even in the significantly worse conditions ($\mu_R \ll \mu_T$) in comparison with the direct measurement at the target system.

Now, we would like to comment on the behavior of $SNR_{DR}$ and $SNR_{GS}$. If we would study their SNR, $SNR_{GS}/SNR_{DR}\approx \sqrt{K}\eta_R/\eta_T$, we can find the dependence only on $K$ the number of trials/modes, i.e., no dependence on the background. However, for the practical application, we are interested in the number of trials/photons $K$. Here we see, that for the same probability of detection, assuming $\eta_R=\eta_T$, we need at the target quadratically more photons than in the QI system. In other words, if we would have 10-times higher background, at the quantum radar side we need to compensate it with 10-times more signal photons, but, at the target, we need 100-more photons to compensate the higher background.

In the rest of this section, we will focus on the more realistic estimation of the factors $\eta_T$ and $\eta_R$ for the quantum radar range $25 - 200$~km. First, the number of background solar photons can be modeled by the black body radiation gives $N_B \approx 10^4$ photons for the temperature at the surface of the sun approx. $T=6000^\circ$~K.

The losses due to the propagation of photons in the atmosphere were described in details in \cite{Lanzagorta:QuantumRadar,lanzagorta_low-brightness_2015}. The propagation of individual photons that includes scattering and absorption can be modeled by the chain of beam splitters \cite{Loudon:optics,loudon_quantum_2000}. As a non-trivial result, we can use the same attenuation factors as are known for the classical radars \cite{skolnik_radar_2008}. We estimated the attenuation factors for good weather (visibility 300~m) at $15^\circ$C to $\eta_{A, r=25km}=0.98$ and $\eta_{A, r=200km}=0.82$. For bad weather (visibility 30~m) at the same temperature we have $\eta_{A, r=25km}=0.50$ and $\eta_{A, r=200km}=0.004$.

The factor $\eta_{X}$ we will guess as
\begin{equation}\label{}
  \eta_{X} = \frac{\langle I_s \rangle}{\langle I_i \rangle} = \frac{\sigma_Q}{4\pi R^2},
\end{equation}
where $\langle I_i \rangle$ is the photon intensity reaching the target and $\langle I_s \rangle$ is the photon intensity reaching back the quantum radar. $\sigma_Q$ is the quantum radar cross section (RCS) and the factor $1/4\pi R^2$ correspond to the reflected spherical wave at large distance $R$.
Primary, note that for quantum radar we can employ the quantum version of the radar equation \cite{Lanzagorta:QuantumRadar,Liu_radar-equation}. First, we need to guess the quantum radar cross section (QRCS), where are some published works on this topic e.g., \cite{lanzagorta_quantum_2010,Liu_radar-equation,brandsema_theoretical_2016}. Since the reliable calculation of $\sigma_Q$ of macroscopic object is very challenging, we apply only the classical RCS that is the limit case for number of photons in the signal beam $N_S \rightarrow \infty$ in the quantum RCS. For the typical jet fighter the classical cross section is approximately $\sigma_C = 2$~m$^2$. These considerations yield $\eta_{X,r=25km}\approx 2.5\times10^{-10}$ and $\eta_{X,r=200km}\approx 4.0\times10^{-12}$.

The factor $\eta_{DA} = 0.005$ we set as the effective area of the detector is $10\times10~$cm of the radar cross section (RCS) assumed as $\sigma_C = 2$~m$^2$. All the guessed factors lead to the $\eta_R / \eta_T$ ratio of orders $\sim10^{-8} - 10^{-12}$. This means that in case of the SP schema, the typical values of $R_M$ are $R_M \approx 10^{-6} - <10^{-30}$ which means definite advantage for the QRWR for the considered conditions. Note, we omit other factors such as gains, and other characteristics of transmitter and receivers. We generally assume only the detection efficiency $\eta_{DE}=0.5$.

\begin{figure}[htb]
\centering
\includegraphics[width=0.8\textwidth]{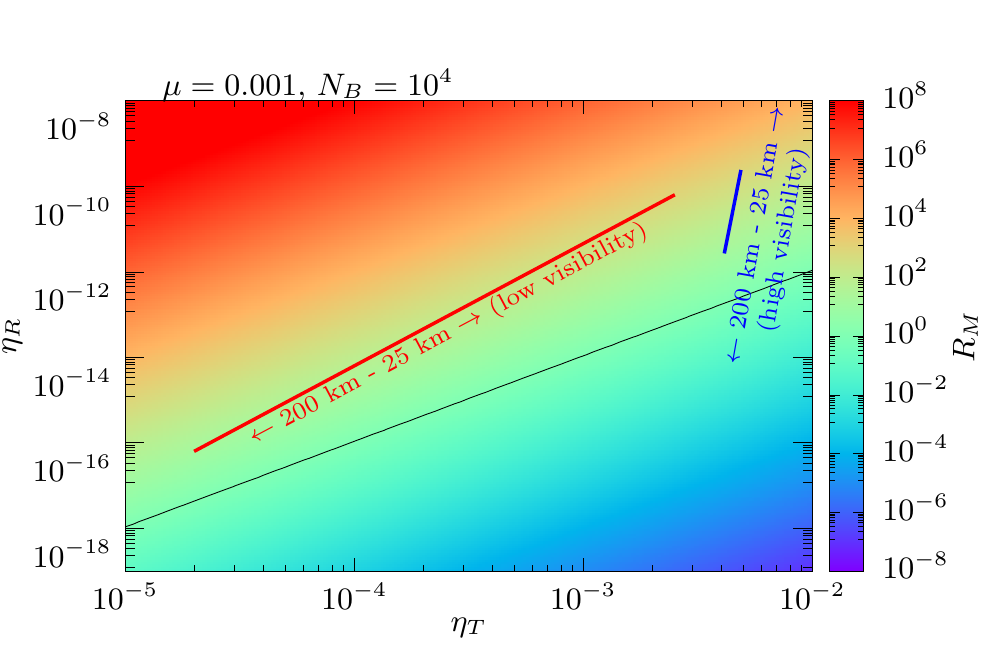}
\caption{Ratio $R_M$ as a function of $\eta_R$ and $\eta_T$ for fixed number of background photons $N_B=10^4$. The black line visualize situation for $R_M=1$. The thick red and blue lines represent the situation for qauntum radar range $25-200$~km for bad and good weather, respective.}
\label{fig:GS-1}
\end{figure}

In the case of the GS schema, the results are represented in Fig.~\ref{fig:GS-1} calculated according \eqref{eq:SNR-DR} and \cite{Guha:2009iie,Barzanjeh:2015zpa}. The thin black line shows the situation for $R_M=1$. The thick red and blue lines represent our estimation for parameters $\eta_R$ and $\eta_T$ for the range $25-200$~km described above for bad (low visibility) and good (high visibility) weather, respective. In both cases, we see that we are in the region of $R_M>1$ that means the advantage of the quantum radar system.
This demonstrates the benefit of fully utilizing of quantum entanglement, which gives strong advantage even for very difficult condition as $\eta_T \gg \eta_R$.

Also, it is suitable to note that the proposed QRWR protocol based on the direct measurement is universal and independent on the quantum radar protocol. There are other quantum radar protocols such as \cite{2010SPIE.7702E..0IS,2019arXiv190300101L,2019arXiv190511939P,2019arXiv190502672M}. However, most of them will be at the level between the considered SP and GS schema of the quantum illumination.

Nevertheless, we studied only the very essence of the quantum illumination principle with a strong assumption on the background. Besides the complex situation with the background, there are many more questions and challenges at the technical level. For example, the target system needs to know which frequencies (which modes) are used for the signal beam, otherwise will need to spend time on the scanning of more frequencies.
This can be especially difficult in the application at radio frequencies, particularly at microwave frequencies bands with increasing spectrum demand and using \cite{RFdemands,brown_history_1984}.
The significant role will also play the detectors on both sides, their efficiency, size, operating parameters, etc.
All these factors were not assumed in this study and require a more detailed review, especially the detection strategy in the environment with more source of different background and noise photons.

%
%
\section{Conclusions}\label{sec:concl}

In this report, we studied the possibility of the detection of QI measurement that tries to detect the target. We discuss how such detection could work, and we highlighted that the main problem of this detection lies in the possibility of reliable detect and separate background photons. This appears to be a challenge in the microwave regimes where, for example, the number of solar background photons is orders of magnitude more than in the optical regime besides the other sources of microwave photons such as human-made microwave transmissions or cosmic background.

The final decision about the QRWR feasibility will require more theoretical as well as experimental studies of the microwave background.

In the next part, we assumed a well-controlled thermal background, and we studied what chances to detect the quantum illumination based radar at the target are. Our presumption is based on the fact of different losses, where we expect significantly smaller losses at the target than in the radar, $\eta_T \gg \eta_R$. For the QI system, we assumed two possible realizations, the single-photon (SP) approach using simple post-measurement correlations. The second approach is based on the Gaussian states (GS) realized by the two-mode squeezing states and homodyne detection receiver, which utilizes the full entanglement advantage.

We found a very good chance to detect the presence of the QI measurement in the case of the SP method for QI, even with high probability that the QI system could be discovered before the target itself will be exposed. This gives to the target time for some counter action such as the change of trajectory.

In the case of the GS approach for the QI, the question now is not about how much in advance the target will detect the QI system before itself will be discovered, but if even the QI system presence will be detected. GS method is an example of the direct utilization of the quantum entanglement, which gives a considerable advantage. Nevertheless, this method has a substantial restriction in the precedent knowledge of the target for the simultaneous joint measurement of the signal and ancilla beams.

Finally, we guessed realistic values of factors $\eta_R$ and $\eta_T$ in the microwave regimes for the quantum radar range $25 - 200$~km. In the case of the SP schema, we confirmed the considerable advantage of the QRWR. However, in the case of GS schema, the quantum radar will have still advantage; however, it could be detected by the proposed QRWR under the considered background.

%
%
\bibliographystyle{utphys}
\addcontentsline{toc}{section}{References}
\bibliography{refs}

\providecommand{\href}[2]{#2}\begingroup\raggedright\begin{thebibliography}{10}

\bibitem{Lloyd:2008lqd}
S.~Lloyd, ``Enhanced sensitivity of photodetection via quantum illumination,''
  \href{http://dx.doi.org/10.1126/science.1160627}{{\em Science} {\bfseries
  321} no.~5895, (2008) 1463--1465}.

\bibitem{Shapiro:2009qiv}
J.~H. Shapiro and S.~Lloyd, ``Quantum illumination versus coherent-state target
  detection,'' {\em New Journal of Physics} {\bfseries 11} no.~6, (2009)
  063045, \href{http://arxiv.org/abs/0902.0986}{{\ttfamily arXiv:0902.0986
  [quant-ph]}}.

\bibitem{Tan:2008qig}
S.-H. Tan, B.~I. Erkmen, V.~Giovannetti, S.~Guha, S.~Lloyd, L.~Maccone,
  S.~Pirandola, and J.~H. Shapiro, ``Quantum illumination with gaussian
  states,'' \href{http://dx.doi.org/10.1103/PhysRevLett.101.253601}{{\em Phys.
  Rev. Lett.} {\bfseries 101} (Dec, 2008) 253601}.

\bibitem{Lopaeva:2013urt}
E.~D. Lopaeva, I.~Ruo~Berchera, I.~P. Degiovanni, S.~Olivares, G.~Brida, and
  M.~Genovese, ``Experimental realization of quantum illumination,''
  \href{http://dx.doi.org/10.1103/PhysRevLett.110.153603}{{\em Phys. Rev.
  Lett.} {\bfseries 110} (Apr, 2013) 153603},
  \href{http://arxiv.org/abs/1303.4304}{{\ttfamily arXiv:1303.4304
  [quant-ph]}}.

\bibitem{Zhang:2015kjr}
Z.~Zhang, S.~Mouradian, F.~N.~C. Wong, and J.~H. Shapiro,
  ``Entanglement-enhanced sensing in a lossy and noisy environment,''
  \href{http://dx.doi.org/10.1103/PhysRevLett.114.110506}{{\em Phys. Rev.
  Lett.} {\bfseries 114} (Mar, 2015) 110506}.

\bibitem{Guha:2009iie}
S.~Guha and B.~I. Erkmen, ``Gaussian-state quantum-illumination receivers for
  target detection,'' \href{http://dx.doi.org/10.1103/PhysRevA.80.052310}{{\em
  Phys. Rev. A} {\bfseries 80} no.~5, (2009) 052310},
  \href{http://arxiv.org/abs/0911.0950}{{\ttfamily arXiv:0911.0950
  [quant-ph]}}.

\bibitem{Invernizzi:2011fwa}
C.~Invernizzi, M.~G.~A. Paris, and S.~Pirandola, ``Optimal detection of losses
  by thermal probes,'' \href{http://dx.doi.org/10.1103/PhysRevA.84.022334}{{\em
  Phys. Rev. A} {\bfseries 84} (Aug, 2011) 022334},
  \href{http://arxiv.org/abs/1011.2785}{{\ttfamily arXiv:1011.2785
  [quant-ph]}}.

\bibitem{Chang:2018nfg}
C.~W. {Sandbo Chang}, A.~M. {Vadiraj}, J.~{Bourassa}, B.~{Balaji}, and C.~M.
  {Wilson}, ``{Quantum-Enhanced Noise Radar},'' {\em arXiv e-prints} (Dec,
  2018) arXiv:1812.03778, \href{http://arxiv.org/abs/1812.03778}{{\ttfamily
  arXiv:1812.03778 [quant-ph]}}.

\bibitem{Lanzagorta:QuantumRadar}
M.~Lanzagorta, \href{http://dx.doi.org/10.2200/S00384ED1V01Y201110QMC005}{{\em
  Quantum Radar}}.
\newblock Synthesis Lectures on Quantum Computing. Morgan {\&} Claypool
  Publishers, 2011.
\newblock \url{http://dx.doi.org/10.2200/S00384ED1V01Y201110QMC005}.

\bibitem{Barzanjeh:2015zpa}
S.~Barzanjeh, S.~Guha, C.~Weedbrook, D.~Vitali, J.~H. Shapiro, and
  S.~Pirandola, ``Microwave quantum illumination,''
  \href{http://dx.doi.org/10.1103/PhysRevLett.114.080503}{{\em Phys. Rev.
  Lett.} {\bfseries 114} no.~8, (2015) 080503},
\href{http://arxiv.org/abs/1503.00189}{{\ttfamily arXiv:1503.00189
  [quant-ph]}}.

\bibitem{England:2019nfj}
D.~G. {England}, B.~{Balaji}, and B.~J. {Sussman}, ``{Quantum-enhanced standoff
  detection using correlated photon pairs},''
  \href{http://dx.doi.org/10.1103/PhysRevA.99.023828}{{\em Physical Review A}
  {\bfseries 99} no.~2, (Feb, 2019) 023828},
  \href{http://arxiv.org/abs/1811.04113}{{\ttfamily arXiv:1811.04113
  [quant-ph]}}.

\bibitem{Pirandola:2018dfr}
S.~{Pirandola}, B.~R. {Bardhan}, T.~{Gehring}, C.~{Weedbrook}, and S.~{Lloyd},
  ``{Advances in photonic quantum sensing},''
  \href{http://dx.doi.org/10.1038/s41566-018-0301-6}{{\em Nature Photonics}
  {\bfseries 12} (Dec, 2018) 724--733},
  \href{http://arxiv.org/abs/1811.01969}{{\ttfamily arXiv:1811.01969
  [quant-ph]}}.

\bibitem{Genovese:2016rjb}
M.~Genovese, ``Real applications of quantum imaging,''
  \href{http://dx.doi.org/10.1088/2040-8978/18/7/073002}{{\em Journal of
  Optics} {\bfseries 18} no.~7, (2016) 073002},
  \href{http://arxiv.org/abs/1601.06066}{{\ttfamily arXiv:1601.06066
  [quant-ph]}}.

\bibitem{Meda:2017uir}
A.~Meda, E.~Losero, N.~Samantaray, F.~Scafirimuto, S.~Pradyumna, A.~Avella,
  I.~Ruo-Berchera, and M.~Genovese, ``Photon-number correlation for quantum
  enhanced imaging and sensing,''
  \href{http://dx.doi.org/10.1088/2040-8986/aa7b27}{{\em Journal of Optics}
  {\bfseries 19} no.~9, (2017) 094002},
  \href{http://arxiv.org/abs/1612.08103}{{\ttfamily arXiv:1612.08103
  [quant-ph]}}.

\bibitem{buttler_daylight_2000}
W.~T. Buttler, R.~J. Hughes, S.~K. Lamoreaux, G.~L. Morgan, J.~E. Nordholt, and
  C.~G. Peterson, ``Daylight {Quantum} {Key} {Distribution} over 1.6 km,''
  \href{http://dx.doi.org/10.1103/PhysRevLett.84.5652}{{\em Physical Review
  Letters} {\bfseries 84} no.~24, (June, 2000) 5652--5655}.

\bibitem{nordholt_present_2002}
J.~E. Nordholt, R.~J. Hughes, G.~L. Morgan, C.~G. Peterson, and C.~C. Wipf,
  \href{http://dx.doi.org/10.1117/12.464085}{``Present and future free-space
  quantum key distribution,''} in {\em Free-{Space} {Laser} {Communication}
  {Technologies} {XIV}}, vol.~4635, pp.~116--126.
\newblock International Society for Optics and Photonics, Apr., 2002.

\bibitem{erven_entangled_2008}
C.~Erven, C.~Couteau, R.~Laflamme, and G.~Weihs, ``Entangled quantum key
  distribution over two free-space optical links,''
  \href{http://dx.doi.org/10.1364/OE.16.016840}{{\em Optics Express} {\bfseries
  16} no.~21, (Oct., 2008) 16840--16853}.

\bibitem{bourgoin_comprehensive_2013}
J.-P. Bourgoin, E.~Meyer-Scott, B.~L. Higgins, B.~Helou, C.~Erven, H.~Hübel,
  B.~Kumar, D.~Hudson, I.~D.~Souza, R.~Girard, R.~Laflamme, and T.~Jennewein,
  ``A comprehensive design and performance analysis of low {Earth} orbit
  satellite quantum communication,''
  \href{http://dx.doi.org/10.1088/1367-2630/15/2/023006}{{\em New Journal of
  Physics} {\bfseries 15} no.~2, (Feb., 2013) 023006}.

\bibitem{er-long_background_2005}
M.~Er-long, H.~Zheng-fu, G.~Shun-sheng, Z.~Tao, D.~Da-sheng, and G.~Guang-can,
  ``Background noise of satellite-to-ground quantum key distribution,''
  \href{http://dx.doi.org/10.1088/1367-2630/7/1/215}{{\em New Journal of
  Physics} {\bfseries 7} (Oct., 2005) 215--215}.

\bibitem{weedbrook_quantum_2010}
C.~Weedbrook, S.~Pirandola, S.~Lloyd, and T.~C. Ralph, ``Quantum {Cryptography}
  {Approaching} the {Classical} {Limit},''
  \href{http://dx.doi.org/10.1103/PhysRevLett.105.110501}{{\em Physical Review
  Letters} {\bfseries 105} no.~11, (Sept., 2010) 110501}.

\bibitem{weedbrook_continuous-variable_2012}
C.~Weedbrook, S.~Pirandola, and T.~C. Ralph, ``Continuous-variable quantum key
  distribution using thermal states,''
  \href{http://dx.doi.org/10.1103/PhysRevA.86.022318}{{\em Physical Review A}
  {\bfseries 86} no.~2, (Aug., 2012) 022318}.

\bibitem{RFdemands}
J.~{Reed}, M.~{Vassiliou}, and S.~{Shah}, ``The role of new technologies in
  solving the spectrum shortage [point of view],''
  \href{http://dx.doi.org/10.1109/JPROC.2016.2562758}{{\em Proceedings of the
  IEEE} {\bfseries 104} no.~6, (June, 2016) 1163--1168}.

\bibitem{brown_history_1984}
W.~C. Brown, ``The {History} of {Power} {Transmission} by {Radio} {Waves},''
  \href{http://dx.doi.org/10.1109/TMTT.1984.1132833}{{\em IEEE Transactions on
  Microwave Theory and Techniques} {\bfseries 32} no.~9, (Sept., 1984)
  1230--1242}.

\bibitem{Lopaeva:2014tvv}
E.~D. Lopaeva, I.~R. Berchera, S.~Olivares, G.~Brida, I.~P. Degiovanni, and
  M.~Genovese, ``A detailed description of the experimental realization of a
  quantum illumination protocol,''
  \href{http://dx.doi.org/10.1088/0031-8949/2014/T160/014026}{{\em Physica
  Scripta} {\bfseries 2014} no.~T160, (2014) 014026},
  \href{http://arxiv.org/abs/1307.3876}{{\ttfamily arXiv:1307.3876
  [quant-ph]}}.

\bibitem{lanzagorta_low-brightness_2015}
M.~Lanzagorta, \href{http://dx.doi.org/10.1117/12.2177577}{``Low-brightness
  quantum radar,''} in {\em Radar {Sensor} {Technology} {XIX}; and {Active} and
  {Passive} {Signatures} {VI}}, vol.~9461, p.~946113.
\newblock International Society for Optics and Photonics, May, 2015.

\bibitem{Loudon:optics}
J.~R. {Jeffers}, N.~{Imoto}, and R.~{Loudon}, ``{Quantum optics of
  traveling-wave attenuators and amplifiers},''
  \href{http://dx.doi.org/10.1103/PhysRevA.47.3346}{{\em Phys.Rev.A} {\bfseries
  47} no.~4, (Apr, 1993) 3346--3359}.

\bibitem{loudon_quantum_2000}
R.~Loudon, {\em The {Quantum} {Theory} of {Light}}.
\newblock OUP Oxford, 2000.

\bibitem{skolnik_radar_2008}
M.~Skolnik, {\em Radar {Handbook}, {Third} {Edition}}.
\newblock Electronics electrical engineering. McGraw-Hill Education, 2008.

\bibitem{Liu_radar-equation}
K.~{Liu}, H.~{Xiao}, H.~{Fan}, and Q.~{Fu}, ``Analysis of quantum radar cross
  section and its influence on target detection performance,''
  \href{http://dx.doi.org/10.1109/LPT.2014.2317759}{{\em IEEE Photonics
  Technology Letters} {\bfseries 26} no.~11, (June, 2014) 1146--1149}.

\bibitem{lanzagorta_quantum_2010}
M.~Lanzagorta, \href{http://dx.doi.org/10.1117/12.854935}{``Quantum radar cross
  sections,''} in {\em Quantum {Optics}}, vol.~7727, p.~77270K.
\newblock International Society for Optics and Photonics, June, 2010.

\bibitem{brandsema_theoretical_2016}
M.~J. Brandsema, R.~M. Narayanan, and M.~Lanzagorta, ``Theoretical and
  computational analysis of the quantum radar cross section for simple
  geometrical targets,''
  \href{http://dx.doi.org/10.1007/s11128-016-1494-6}{{\em Quantum Information
  Processing} {\bfseries 16} no.~1, (Dec., 2016) 32}.

\bibitem{2010SPIE.7702E..0IS}
I.~{Smith}, James~F., \href{http://dx.doi.org/10.1117/12.849810}{``{Quantum
  interferometer and radar theory based on N00N, M and M or linear combinations
  of entangled states},''} in {\em Quantum Information and Computation VIII},
  vol.~7702 of {\em Society of Photo-Optical Instrumentation Engineers (SPIE)
  Conference Series}, p.~77020I.
\newblock Apr, 2010.

\bibitem{2019arXiv190300101L}
D.~{Luong}, C.~W. {Sandbo Chang}, A.~M. {Vadiraj}, A.~{Damini}, C.~M. {Wilson},
  and B.~{Balaji}, ``{Receiver Operating Characteristics for a Prototype
  Quantum Two-Mode Squeezing Radar},'' {\em arXiv e-prints} (Feb, 2019)
  arXiv:1903.00101, \href{http://arxiv.org/abs/1903.00101}{{\ttfamily
  arXiv:1903.00101 [quant-ph]}}.

\bibitem{2019arXiv190511939P}
I.~{Peshko}, D.~{Mogilevtsev}, I.~{Karuseichyk}, A.~{Mikhalychev}, A.~P.
  {Nizovtsev}, G.~Y. {Slepyan}, and A.~{Boag}, ``{Quantum noise radar:
  superresolution with quantum antennas by accessing spatiotemporal
  correlations},'' {\em arXiv e-prints} (May, 2019) arXiv:1905.11939,
  \href{http://arxiv.org/abs/1905.11939}{{\ttfamily arXiv:1905.11939
  [quant-ph]}}.

\bibitem{2019arXiv190502672M}
L.~{Maccone} and C.~{Ren}, ``{Quantum radar},'' {\em arXiv e-prints} (May,
  2019) arXiv:1905.02672, \href{http://arxiv.org/abs/1905.02672}{{\ttfamily
  arXiv:1905.02672 [quant-ph]}}.

\end{thebibliography}\endgroup

\end{document}